%% file: main.tex
\documentclass[runningheads]{llncs}

\usepackage[T1]{fontenc}
\usepackage{graphicx}
\usepackage{cite}
\usepackage{amsmath,amssymb,amsfonts}
\usepackage{algorithmic}
\usepackage{graphicx}
\usepackage{textcomp}
\def\BibTeX{{\rm B\kern-.05em{\sc i\kern-.025em b}\kern-.08em
    T\kern-.1667em\lower.7ex\hbox{E}\kern-.125emX}}
\usepackage{soul}
\usepackage{xcolor}
\usepackage{xspace}
\usepackage[table,xcdraw]{xcolor}
\usepackage{threeparttable}
\usepackage{hyperref}
\usepackage{multirow}
\usepackage{multicol}
\usepackage{todonotes}
\usepackage{enumitem}
\usepackage[most]{tcolorbox}
\usepackage{orcidlink}

\newcommand{\tool}{\textsc{LexTester}\xspace}
\newcommand{\botium}{\textsc{Botium}\xspace}
\newcommand{\numChatbots}{five\xspace}

\newtcolorbox{shadedbox}{
drop shadow southeast,
breakable,
enhanced jigsaw,
colback=white,
boxrule=0.80pt,
left=0.3em,
right=0.3em,
top=0.1em,
bottom=0.05em
}

\begin{document}
\title{A Model-based Testing Technique for Amazon Lex Task-based Chatbots}
\titlerunning{A Model-based Testing Technique for Amazon Lex Task-based Chatbots}

\author{
Diego Clerissi\thanks{Corresponding Author}\orcidlink{0000-0001-7651-0400} \and
Alessandro Vasina \and 
Leonardo Mariani\orcidlink{0000-0001-9527-7042}}

\authorrunning{D. Clerissi et al.}

\institute{Department of Informatics, Systems and Communication \\ University of Milano-Bicocca \\
Milan, Italy\\
\email{\{firstname\}.\{lastname\}@unimib.it}}

\maketitle

\begin{abstract}

Task-based chatbots are nowadays widely adopted software systems, usually integrated into real-world applications and communication channels, designed to assist users in completing tasks through conversational interfaces. 
Like any other software, even chatbots are prone to bugs. Despite their increasing pervasiveness in everyday activities, existing techniques for assessing their quality still exhibit several limitations, such as the simplicity of generated test scenarios and oracle weaknesses. 

In this paper, we present \tool, an automated model-based testing technique for Amazon Lex chatbots. The technique explores the conversational space of the chatbot under test to generate a Dialog Graph of all possible interactions, from which an executable test suite is generated according to different coverage strategies. \tool was evaluated against the state-of-the-practice chatbot testing tool Botium on \numChatbots Amazon Lex chatbots, 
consistently outperforming it in all subjects, generating more tests with nearly double complexity, achieving overall 83-95\% coverage of conversational elements, and improving fault detection effectiveness by up to four times at comparable time costs.   
\keywords{Task-based Chatbot \and Model-based Testing \and Amazon Lex}
\end{abstract}

\section{Introduction}\label{sec:introduction}\input{introduction}
\section{Background} \label{sec:background}\input{background}

\section{\tool}\label{sec:tool}\input{tool}

\section{Empirical Evaluation}\label{sec:evaluation}\input{evaluation}
\section{Related Work}\label{sec:related-work}\input{related-work}
\section{Conclusion}\label{sec:conclusion}\input{conclusion}

\begin{credits}
\subsubsection{\discintname}
The authors have no competing interests to declare that are relevant to the content of this article.
\end{credits}

\bibliographystyle{splncs04}
\bibliography{references}

\end{document}

%% file: introduction.tex
In recent years, \emph{task-based chatbots}, i.e., software designed to deliver functionalities through conversations\cite{grudin2019chatbots,adamopoulou2020overview,adamopoulou2020chatbots}
, have gained popularity due to their integration into real-world applications. Advancements in technology have made them ubiquitous in a multitude of domains, including e-commerce, healthcare, and customer help-desk~\cite{cui2017superagent,fiore2019forgot,de2021s} and a large number of design platforms have emerged, with Dialogflow, Amazon Lex, Rasa, Microsoft Bot, and IBM Watson being systematically acknowledged as the most popular platforms~\cite{perez2021choosing,abdellatif2021comparison,motger2022software,benaddi2024systematic}
, in both open-source and commercial settings.

Given chatbot pervasiveness, tailored approaches to ensure their quality have recently emerged, yet raising novel challenges~\cite{cabot2021testing,deriu2021survey,li2022review,lambiase2024motivations}.
Unlike traditional testing approaches that validate the software, for instance, based on API calls or GUI actions, chatbots combine conventional software layers managing business logic with conversational interfaces that must interpret user requests through natural-language interactions, whose responses must also be interpreted to determine whether a task was correctly performed, raising issues in terms of oracles.

Initial efforts in chatbot quality assurance targeted the speech recognition module~\cite{iwama2019automated,asyrofi2020crossasr} 
and non-functional aspects~\cite{chatbottest}
. 
Later, testing chatbots as a whole was addressed by employing input mutation for robustness testing~\cite{ruane2018botest,guichard2019assessing,liu2021dialtest} and metamorphic testing~\cite{chen2018metamorphic} to address the oracle problem~\cite{bozic2019testing,bovzic2022ontology}. However, these proposals demand extensive manual intervention and deep prior knowledge of the chatbot behavior. 

\botium is the state-of-the-practice framework for automated end-to-end task-based chatbot testing~\cite{botium}, originally developed by Botium GmbH (now Cyara). \botium provides connectors to several widely used chatbot design platforms and frameworks. Subsequent work has exploited \botium to achieve advances in testing, particularly in test diversity~\cite{bravo2020testing,rapisarda2025test} and improved test coverage~\cite{canizares2024coverage}.
Still, all these tools present limitations in test effectiveness, as tests by construction are limited to single request-response interactions,  produce only regression oracles, and are susceptible to flakiness\footnote{\emph{``A flaky test is a test that passes and fails periodically without any code change''}~\cite{zheng2021research}}~\cite{ferdinando2024mutabot,rapisarda2025test,clerissi2025towards,masserini2025brasato,masserini2026assessing}.
Further, all the tools extending \botium only address Dialogflow and Rasa as chatbot development platforms, thus neglecting Amazon Lex despite being one of the most popular commercial solutions~\cite{perez2021choosing,benaddi2024systematic}.  

To address such a need, in this paper, we present \tool, an automated model-based testing approach for Amazon Lex task-based chatbots. We empirically compared \tool with \botium, evaluating the complexity of the generated tests, conversation coverage capabilities, fault detection, and test generation efficiency. To the best of our knowledge, \botium is the only other tool specifically designed for Amazon Lex chatbots. 

The paper provides the following contributions: 
\begin{enumerate}[leftmargin=*, nosep]
    \item It proposes \tool, the first model-based testing technique for Amazon Lex task-based chatbots. \tool first derives a \textit{Dialog Graph} representing all possible conversations of a chatbot, and then generates executable tests that exercise such conversations according to a coverage criterion;
    \item It presents empirical results collected from prebuilt and third-party chatbots that show how \tool outperforms \botium in terms of test complexity, conversation coverage, and fault detection, with comparable time efficiency;
    \item It makes available a replication package with references to the \tool source code, usage instructions, and results for experimental adaptation and replication, found at: \url{https://gitlab.com/lextester1/lextester-exp}.
\end{enumerate}

The paper is organized as follows. Section~\ref{sec:background} introduces the key concepts of task-based chatbots and testing with \botium. Section~\ref{sec:tool} describes \tool. Section~\ref{sec:evaluation} presents the experiment, comparing \tool with \botium. Section~\ref{sec:related-work} discusses related work. Finally, Section \ref{sec:conclusion} provides final remarks.

%% file: background.tex
This section introduces some background about the structure and functionality of task-based chatbots, characteristics of Amazon Lex chatbots, and core concepts of conversational testing in \botium.

\subsection{Task-based Chatbots}

\begin{figure*}[t!]
    \centering
\includegraphics[width=0.9\linewidth, trim=0.2cm 8.8cm 9.0cm 0.0cm, clip=true]{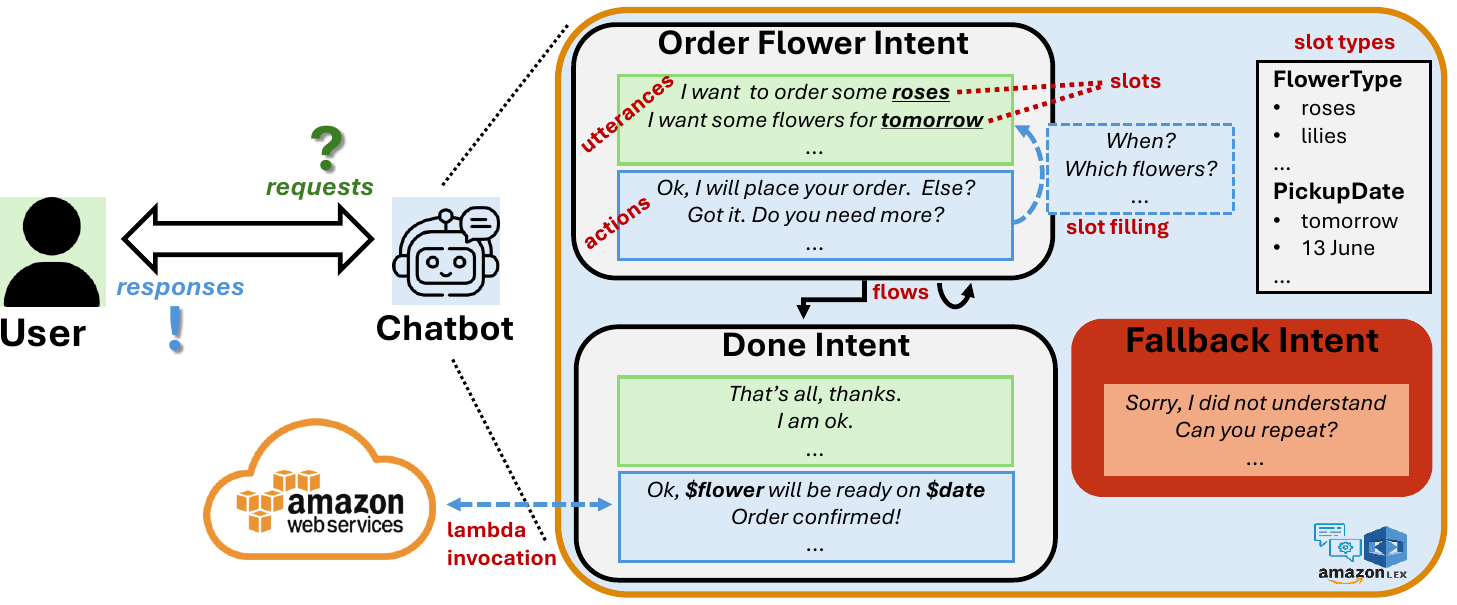}
    \vspace{-3mm}
    \caption{Task-based chatbot architecture in Amazon Lex.}
    \label{fig:chatbot}
    \vspace{-5mm}
\end{figure*}

\emph{Task-based chatbots} are software systems designed to interact with users through conversations, structured as textual, vocal, or visual interactions, and integrated with backend logic and external services. Their primary goal is to efficiently fulfill user requests, prioritizing task completion over casual chat~\cite{adamopoulou2020overview,adamopoulou2020chatbots}.
Figure \ref{fig:chatbot} captures the key concepts underlying task-based chatbots, as commonly supported by most popular development platforms~\cite{canizares2022automating}, including Amazon Lex. 

A task-based chatbot is designed to handle one or more \textit{intents}, each one representing a specific user goal (e.g., \texttt{Order Flower} in the Figure corresponds to the user intention of buying some flowers at a given date).

Each intent includes a set of \textit{utterances} (also called \textit{training phrases}) and \textit{actions}.
Utterances are used to train the chatbot to recognize user requests associated with a given intent (e.g., ``I want to order some roses'' is an utterance for the \texttt{Order Flower} intent).
\textit{Actions}, on the other hand, define how the chatbot fulfills a request and generates a response. Actions may consist of simple plain-text responses (e.g., ``Ok, I will place your order. Else?'') or more advanced operations involving business logic execution and interaction with external services (e.g., ``Order confirmed!'' for the \texttt{Done} intent that triggers a function to activate an external service).

Chatbots define \textit{slot types} (also called \textit{entities}), which represent the datatypes the chatbot recognizes (e.g., \texttt{FlowerType}). Slot types are instantiated through \textit{slots} (also called \textit{parameters}) extracted from user utterances (e.g., ``roses'' for the \texttt{FlowerType} slot type in the first utterance of the \texttt{Order Flower} intent) and can be referenced within actions (e.g., \texttt{\$flower} in the first action of the \texttt{Done} intent). When necessary, a \textit{slot filling} mechanism can be employed by the chatbot to prompt the user for missing required slots (e.g., ``Which flowers?'').

Intents are connected to form \textit{flows}, representing the possible conversational paths that interleave user requests and chatbot responses. Additionally, a \textit{fallback} intent can be defined to handle cases where the chatbot fails to understand a user request, prompting the user to rephrase it.

A typical interaction with a chatbot begins with a user request. The chatbot leverages Natural Language Processing (NLP) models trained on utterances to map the request to an intent. Based on the identified intent, the chatbot generates a response and executes any associated actions, passing the extracted slots if required. If no intent is recognized, the fallback intent is triggered. This process continues until the user goal is fulfilled.

\subsection{Amazon Lex Chatbots}

Amazon Lex is the chatbot design platform developed by Amazon (current version V2\footnote{\url{https://docs.aws.amazon.com/lexv2/latest/dg/what-is.html}}), integrated within Amazon Web Services (AWS), natively supporting connection to, among the others, Amazon cloud services, advanced Natural Language Understanding (NLU) technologies, and lambda functions (i.e., serverless computing services that enable the execution of backend logic, see Figure \ref{fig:chatbot}). 

From a structural perspective, an Amazon Lex chatbot adheres to the typical task-based chatbot structure shown in Figure \ref{fig:chatbot}, hence being composed of intents, slot types, user utterances to train the NLU model, and so on.
In addition, Amazon Lex provides a Visual Conversational Builder\footnote{\url{https://docs.aws.amazon.com/lexv2/latest/dg/visual-conversation-builder.html}} tool to design and visualize the user-bot conversational flows by means of interconnected blocks that represent the different units composing a conversation. 

\begin{figure}[h!]
    \centering
    \includegraphics[width=0.8\linewidth]{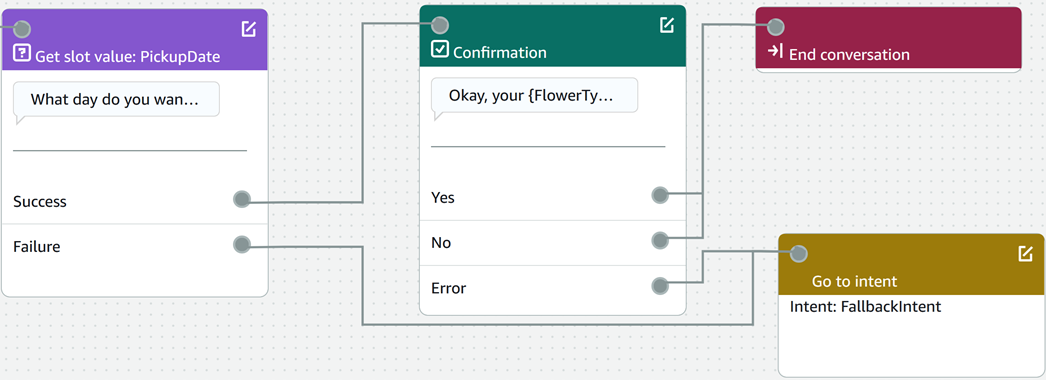}
    \vspace{-3mm}
    \caption{A portion of a conversation modeled in the Visual Conversation Builder.}
    \label{fig:visual-builder}
    \vspace{-1mm}
\end{figure}

Examples of blocks are the \textit{Start} and \textit{End} blocks that initiate and terminate a conversation, \textit{Get slot value} blocks aimed at eliciting slots, \textit{Confirmation} blocks to ask for user confirmation before completing an action, \textit{Code hook} blocks to invoke lambda functions, and \textit{Go to intent} blocks redirecting the conversation flow toward another intent. Figure \ref{fig:visual-builder} shows a portion of a conversation modeled using the Visual Conversation Builder.
Since the Visual Conversation Builder does not support direct export for model-based testing purposes, we took inspiration from it to design our Dialog Graph and integrate it within \tool.

\subsection{\botium Test Cases}
\botium~\cite{botium} is an automated quality assurance framework developed by Botium GmbH in 2018, widely used in both industrial and academic contexts~\cite{bravo2020testing,perez2021choosing,li2022review}. It supports test case generation and execution for various chatbot design platforms, including Dialogflow~\cite{dialogflow}, Amazon Lex~\cite{amazon-lex} and Rasa\cite{rasa}, representing the state-of-the-practice approach for chatbot testing.  

\begin{figure}[t!]
    \centering
    \includegraphics[width=0.35\linewidth, trim=0.0cm 8.3cm 26.0cm 0.0cm, clip=true]{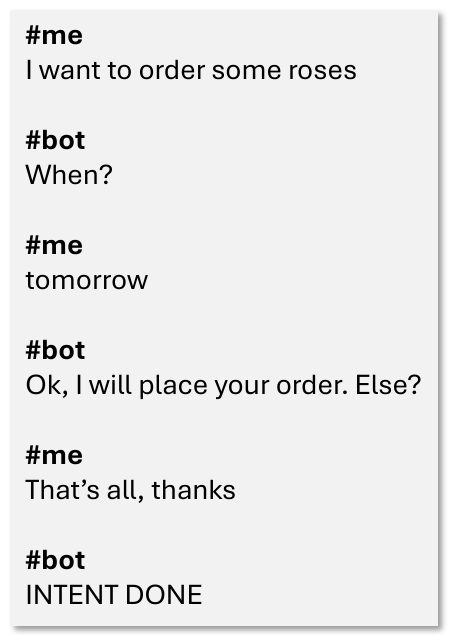}
    \vspace{-3mm}
    \caption{An exemplified \botium test case.}
    \vspace{-6mm}
    \label{fig:test}
\end{figure}

In \botium, a test case is a conversational scenario, composed of text-based steps between the user (annotated with \textbf{\#me}) and the chatbot (annotated with \textbf{\#bot}), derived from the files of the chatbot implementation that encode the possible requests and responses. Figure~\ref{fig:test} shows an exemplified \botium test. When a test is executed, \botium connects to the chatbot on the specific platform and proceeds to simulate the user by sending the utterances established for the scenario to replicate, observing whether the chatbot responses are aligned with the expected ones. \botium supports multiple asserters that can be used to this end
. For instance, the text asserter verifies if the expected response under the \textbf{\#bot} tag matches the actual one. By default, \botium applies the intent asserter (keyword \texttt{INTENT} followed by the intent name that must be activated).

%% file: tool.tex
\begin{figure}
    \centering
    \includegraphics[width=0.9\linewidth, trim=0 8.9cm 11.9cm 0, clip]
    {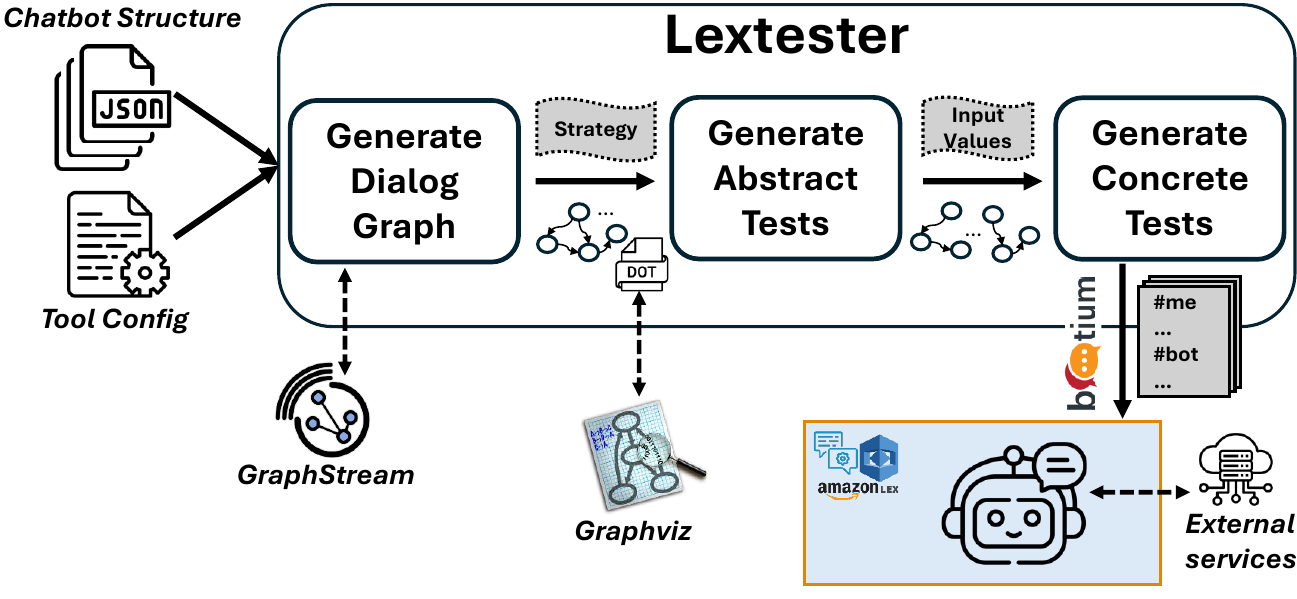}
    \vspace{-4mm}
    \caption{\tool pipeline.}
    \label{fig:tool}
    \vspace{-5mm}
\end{figure}

\tool is a model-based test generator developed in Java for Amazon Lex chatbots. 
The pipeline of the tool is shown in Figure \ref{fig:tool}. 

\tool takes as input the path to a Lex chatbot folder, whose structure is contained in JSON files incorporating the key conversational concepts introduced in Section \ref{sec:background}, and a configuration file defining the coverage strategy for test generation and the data used to generate user input. 
The tool parses the content of the chatbot conversational data, then (i) generates a Dialog Graph, (ii) generates abstract test cases from the graph in the form of conversational paths according to the chosen coverage strategy, and (iii) generates concrete executable test cases in a \botium-like format from abstract test cases, instantiating them with actual input values.
It is worth noting that \tool relies on \botium infrastructure only for test execution.

\subsection{Dialog Graph}
Initially, \tool constructs the Dialog Graph, that is, a directed multigraph inspired by the Amazon Lex Visual Conversation Builder, from the JSON files describing the chatbot structure.
The Dialog Graph is formally defined as $DG = (V, E, type, label)$, where $V$ is the set of vertices representing conversational blocks (e.g., \textit{Start}, \textit{Get slot value}, \textit{Confirmation}, \textit{Go to intent}), $E$ is the set of edges encoding the conversational transitions and outcomes (e.g., success, failure, alternative paths), $type(V)$ identifies the block type associated with vertex $V$, and $label(V)$ stores conversational information associated with $V$, such as utterances or expected slot values.

To generate the graph, \tool iterates over all intents composing the chatbot and analyzes the exported JSON descriptors to infer the conversational block types to create. Depending on the identified type, \tool extracts the relevant conversational information (e.g., slot filling messages, lambda invocations, conditional branches) and associates them with the corresponding vertices and edges. 
Finally, conversational flows are reconstructed by connecting the generated blocks according to the references defined in the chatbot configuration. The Dialog Graph is serialized as a DOT-format file employing GraphStream API\footnote{\url{https://graphstream-project.org/}} for data construction and Graphviz\footnote{\url{https://graphviz.org/}}  for graphical visualization. 

\begin{figure}[t!]
    \centering
    \includegraphics[width=\linewidth]{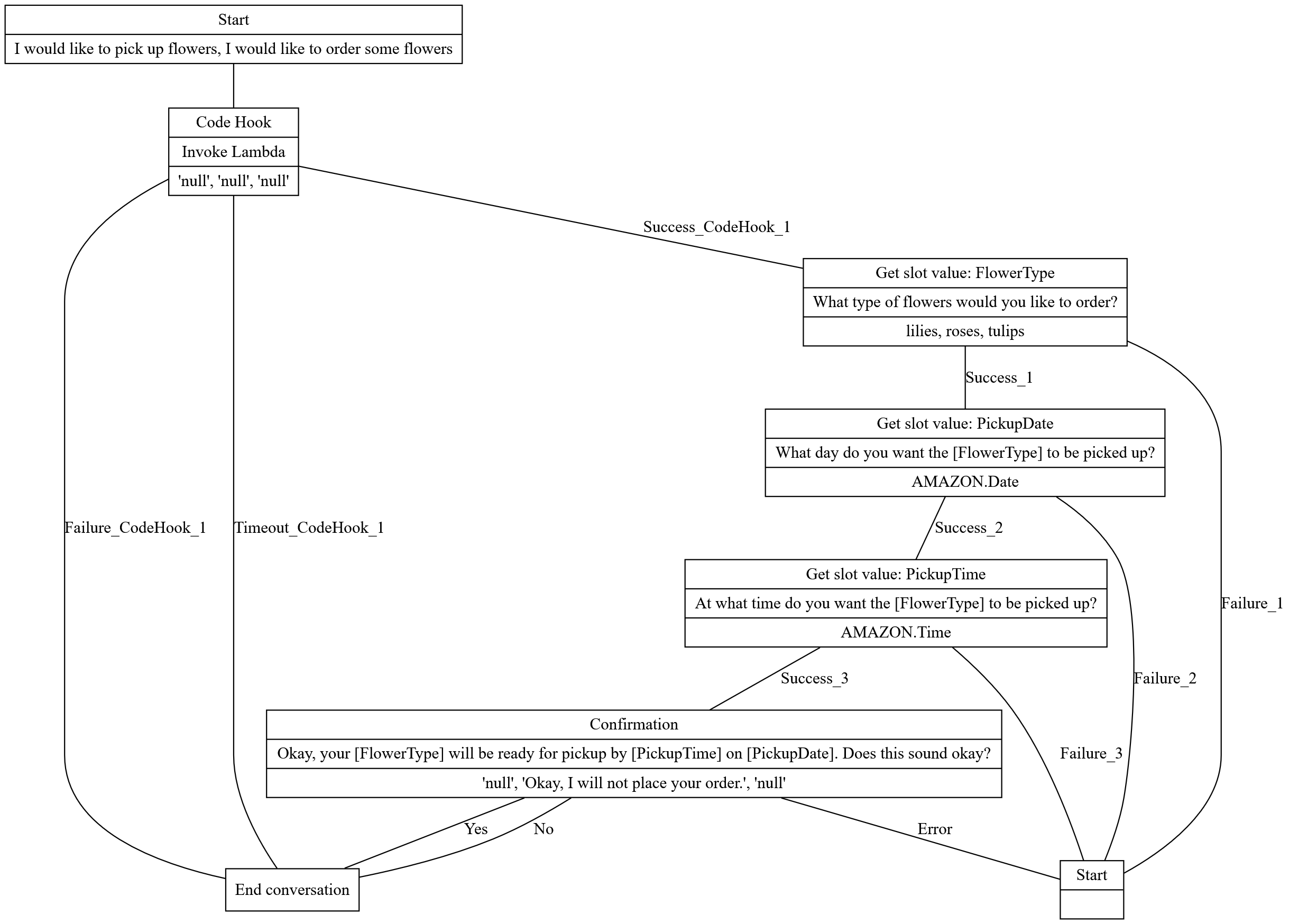}
    \vspace{-3mm}
    \caption{A Dialog Graph generated by \tool.}
    \label{fig:graph}
    \vspace{-4mm}
\end{figure}

Figure \ref{fig:graph} shows an example of a Dialog Graph generated by \tool for a flower-ordering conversation. The flow starts from a \textit{Start} block, followed by a \textit{Code Hook} block invoking a lambda function. The block exposes three possible outcomes (success, failure, and timeout), whose associated values are undefined since they are not statically defined in the chatbot configuration but depend on the lambda implementation. In the success branch, multiple \textit{Get slot value} blocks elicit the required slots (e.g., \texttt{FlowerType}), followed by a \textit{Confirmation} block before reaching the \textit{End conversation} block. Failure branches instead redirect the conversation toward the fallback intent, represented by a separate \textit{Start} block.

\subsection{Abstract Test Cases}
Once the Dialog Graph is built, \tool generates abstract test cases by traversing the graph from the \textit{Start} block toward terminal vertices. \tool is currently implemented to support the N-gram path coverage criterion, i.e., a sequence of up to N consecutive vertices traversed. For example, considering a sequence \textit{Start} $\mapsto$ \textit{Get slot value} $\mapsto$ \textit{Confirmation}, a 3-gram covers the entire conversational subsequence, while a 2-gram covers each pairwise transition independently.
During traversal, each visited vertex contributes conversational information used to incrementally build the test, including user utterances, chatbot responses, and the type of slots expected at that stage of the interaction. 
\tool systematically generates conversational paths that include all distinct N-grams of the graph, thus exercising both short and long interaction patterns.

\subsection{Concrete Test Cases}
Abstract test cases are finally instantiated into executable \botium-like test cases, i.e., sequences of alternating \textbf{\#me} and \textbf{\#bot} interactions. \tool relies on \textit{input values} to fill slots required by the chatbot, which can be either selected from predefined datasets or randomly generated according to the required slot type. During instantiation, placeholders derived from path traversal are replaced with concrete user utterances consistent with the expected slot semantics, producing complete executable conversational tests, adhering to the test structure shown in Figure \ref{fig:test}.

%% file: evaluation.tex
We evaluated \tool against the state-of-the-practice tool \botium on \numChatbots Amazon Lex task-based chatbots, considering the following research questions:
\begin{itemize}
	\item \textbf{RQ1-Test Coverage}: \emph{How effective is \tool in covering conversational elements?} This research question studies the capability of \tool in covering conversational elements compared to \botium; 
    \item \textbf{RQ2-Fault Detection}: \emph{How effective is \tool in detecting faults in conversations?} This research question studies the capability of \tool in detecting faults in conversations compared to \botium; 
    \item \textbf{RQ3-Test Efficiency}: \emph{How does \tool affect test generation efficiency?} This research question studies the cost of applying \tool compared to \botium.
\end{itemize}

To answer \textbf{RQ1}, we measured the percentages of intents, slot types, and slots covered by the executed tests. 

To answer \textbf{RQ2}, we applied mutation testing on the \numChatbots subjects, employing Mutabot state-of-the-art tool~\cite{ferdinando2024mutabot,clerissi2025towards}, measuring mutation score (i.e., the rate of mutants detected over total mutants).

To answer \textbf{RQ3}, we measured the time required by \tool and \botium to generate and execute a test suite, respectively. For \tool, we also included the time required to generate the Dialog Graphs. We evaluated test efficiency in terms of test volume, measured as the number of total tests generated, and test complexity, measured as the minimum, maximum, and average number of user-bot interactions per test.



\subsection{Subject Chatbots}

To investigate the research questions, we selected \numChatbots Lex chatbots: two prebuilt chatbots provided by the Amazon Lex platform and three third-party chatbots available on GitHub, selected considering conversational size (i.e., they present a high number of intents, slot types, and slots), popularity (i.e., they have at least one GitHub star), topic diversity (i.e., they are representative of different domains), and functional complexity (i.e., they implement actual backend logic).

Table~\ref{tab:chatbots} reports the name, topic, and conversational size (i.e., number of intents, slot types, and slots) of each selected chatbot.

\begin{table}[t!]
\caption{Subject Chatbots.}
\vspace{-3mm}
\label{tab:chatbots}
\centering
\scriptsize
\begin{threeparttable}
\resizebox{\columnwidth}{!}{
\begin{tabular}{|>{\raggedright}p{3.1cm}|c|c|c|c|}
\hline
\rowcolor{gray!40}
\textbf{Chatbot} & \textbf{Topic} & \textbf{\# Intents} & \textbf{\# Slot Types} & \textbf{\# Slots} \\
\hline
Air Line Bot\tnote{a} & Travel & 10 & 6 & 13 \\ 
\hline
Book Trip Manager\tnote{b} & Travel & 3 & 2 & 9 \\
\hline
Cloud Assistant\tnote{c} & Business & 12 & 9 & 9 \\
\hline
Movie Recommender\tnote{d} & Entertainment & 2 & 1 & 1 \\
\hline
Order Flowers Bot\tnote{b} & Shopping & 2 & 1 & 3 \\
\hline
\end{tabular}
}
{\fontsize{6}{8}\selectfont
\begin{tablenotes}
    \item[a] \url{https://github.com/aws/lex-helper/tree/main/examples/sample_airline_bot}
    \item[b] \url{https://console.aws.amazon.com} (Prebuilt)
    \item[c] \url{https://github.com/aws-samples/aws-amplify-cloud-assistant-app.git}
    \item[d] \url{https://github.com/cloudacademy/aws-lexv2-chatbot}
\end{tablenotes}
}
\end{threeparttable}
\vspace{-5mm}
\end{table}

\subsection{Experimental Procedure}

To conduct the experiment, we configured each third-party chatbot following the instructions provided in the repositories, while the prebuilt chatbots were generated directly from the platform. 

For \botium, we used the Amazon Lex connector implementation to interact with chatbots on the platform and generate tests. For \tool, we used the implemented N-gram coverage strategy with path length set up to 10.

Coverage 
was measured by inspecting the generated tests and execution outcomes.
For fault injections, we employed the Mutabot tool~\cite{ferdinando2024mutabot,clerissi2025towards}, originally designed for Dialogflow and equipped with 24 operators to mutate chatbot conversational elements, such as removing an intent, changing the name of a slot type, or modifying a user utterance. In this work, we adapted the operators to Amazon Lex, to simulate faults occurring on that platform. Each mutant was inspected, iteratively deployed on the platform, and run against each \botium and \tool test suite, measuring mutation score.
Finally, the time required for test generation and execution was measured using the timers implemented in each tool, while test complexity (i.e., minimum, maximum, and average number of user-bot interactions per test) was measured using an automated script. 

To reduce randomness in the results, we repeated each test execution five times for both \tool and \botium on each chatbot, cleaning chatbot states after each execution (e.g., created resources) to avoid any side effect.

\subsection{Results for RQ1 - Test Coverage}

Table \ref{tab:rq1} shows the test volume and the coverage (\%) of intents, slot types, and slots computed for each chatbot and technique.

\begin{table}[b!]
\caption{Results for RQ1 - Test Volume and Conversational Coverage (\%).}
\vspace{-3mm}
\label{tab:rq1}
\centering
\small
\resizebox{\columnwidth}{!}{
\begin{tabular}{|>{\raggedright}p{3.1cm}|c|c|c|c|c|c|c|c|}
\cline{2-9}
\multicolumn{1}{c|}{} 
& \multicolumn{4}{>{\columncolor{gray!40}}c|}{\textbf{Botium}} 
& \multicolumn{4}{>{\columncolor{gray!40}}c|}{\textbf{LexTester}} \\
\hline
\rowcolor{gray!40}
\textbf{Chatbot}
& \textbf{\# Tests} & \textbf{Intent} & \textbf{Slot Type} & \textbf{Slot}
& \textbf{\# Tests} & \textbf{Intent} & \textbf{Slot Type} & \textbf{Slot} \\
\hline
Air Line Bot & 249 & 90\% & 83\% & 77\% & 2,172 & 90\% & 83\% & \textbf{85\%}  \\
\hline
Book Trip Manager & 85 & 67\% & 40\% & 22\% & 375 & 67\% & \textbf{100\%} & \textbf{100\%}  \\
\hline
Cloud Assistant & 1,133 & 83\% & 89\% & 89\% & 585 & \textbf{92\%} & \textbf{100\%} & \textbf{100\%} \\
\hline
Movie Recommender & 78 & 50\% & 100\% & 100\% & 146 & 50\% & 100\% & 100\% \\
\hline
Order Flowers Bot & 2 & 50\% & 0\% & 0\% & 349 & 50\% & \textbf{100\%} & \textbf{100\%}\\
\hline\hline
\rowcolor{gray!10}
\textbf{Total} 
& 1,547 & 79\% & 73\% & 60\% 
& \textbf{3,627} & \textbf{83\%} & \textbf{95\%} & \textbf{94\%} \\
\hline
\end{tabular}
}
\vspace{-4mm}
\end{table}

In all cases except for the \texttt{Cloud Assistant} chatbot, \tool was able to produce more tests, with test suites 1.9-175 times larger than those produced by \botium. Moreover, \tool achieved the same or higher coverage than \botium in each chatbot.
This is particularly evident for  chatbots of high complexity (e.g., \texttt{Cloud Assistant}).
On the other hand, due to its simplicity, the only chatbot in which both \botium and \tool achieved the same coverage was \texttt{Movie Recommender}.

Concerning intents, none of the tools was able to cover fallback intents, as when no training data is provided for failing scenarios, the tools cannot exploit it for test generation, thus none of them ever achieved 100\% of intent coverage. Overall, \botium and \tool capabilities were very similar, achieving intent coverages of 79\% and 83\%, respectively, with \tool covering one intent missed by \botium, which was unable to use proper data to generate tests. 

The results highlighted better performance in \tool for the coverage of the slot types and the slots. In fact, several chatbots defined slot types and instantiated slots in scenarios based on follow-up interactions (i.e., when a sentence does not provide all required slots, it activates a slot filling mechanism to elicit the missing ones), which \botium was unable to resolve, even in simple chatbots (e.g., in \texttt{Order Flowers Bot} the user is prompted to specify which flower to order, which is never covered by \botium).
Overall, \botium achieved 73\% slot type coverage and 60\% slot coverage, whereas \tool achieved 95\% and 94\%, respectively.


\begin{shadedbox}
    \textbf{Answer to RQ1:} \tool was able to cover more conversational elements than \botium in four cases out of five, particularly for slot types and slots, due to its capability in generating follow-up interactions aimed at stimulating slot filling. 
\end{shadedbox}

\subsection{Results for RQ2 - Fault Detection} 

Table \ref{tab:rq2} shows mutation score computed for faults injected with Mutabot in intents, slot types, and input/output elements pertaining to flows (e.g., a user request, a bot response, or a context variable representing data propagated between intents with a certain lifespan). We inspected the mutants generated to remove equivalent ones, which were not found. Further, we verified that the adapted mutations produced valid  faults for Amazon Lex chatbots. 

\begin{table}[b!]
\caption{Results for RQ2 - Mutation Score (\%).}
\label{tab:rq2}
\vspace{-3mm}
\centering
\resizebox{\columnwidth}{!}{
\begin{tabular}{|>{\raggedright}p{3.1cm}|
c|c|c|
c|c|c|}
\cline{2-7}
\multicolumn{1}{c|}{} 
& \multicolumn{3}{>{\columncolor{gray!40}}c|}{\textbf{Botium}} 
& \multicolumn{3}{>{\columncolor{gray!40}}c|}{\textbf{LexTester}} \\
\hline
\rowcolor{gray!40}
\textbf{Chatbot}
& \textbf{Intent} & \textbf{Slot Type} & \textbf{I/O Flow}
& \textbf{Intent} & \textbf{Slot Type} & \textbf{I/O Flow} \\
\hline

Air Line Bot 
& 18/20 (90\%) & 0/77 (0\%) & 0/2 (0\%)
& 18/20 (90\%) & \textbf{44/77 (57\%)} & \textbf{2/2 (100\%)} \\
\hline

Book Trip Manager 
& 4/6 (67\%) & 0/4 (0\%) & 0/13 (0\%) 
& 4/6 (67\%) & \textbf{4/4 (100\%)} & \textbf{10/13 (77\%)} \\
\hline

Cloud Assistant 
& 20/24 (83\%) & 19/56 (34\%) & 20/48 (42\%)
& \textbf{22/24 (92\%)} & \textbf{28/56 (50\%)} & \textbf{21/48 (44\%)} \\
\hline

Movie Recommender 
& 2/4 (50\%) & 0/4 (0\%) & 18/21 (86\%) 
& 2/4 (50\%) & \textbf{4/4 (100\%)} & \textbf{21/21 (100\%)} \\
\hline

Order Flowers Bot 
& 2/4 (50\%) & 0/3 (0\%) & 0/5 (0\%) 
& 2/4 (50\%) & \textbf{3/3 (100\%)} & \textbf{4/5 (80\%)} \\
\hline\hline

\rowcolor{gray!10}
\textbf{Total} 
& 46/58 (79\%) & 19/144 (13\%) & 38/89 (43\%)
& \textbf{48/58 (83\%)} & \textbf{83/144 (58\%)} & \textbf{58/89 (65\%)} \\
\hline

\end{tabular}
}
\vspace{-6mm}
\end{table}

Considering intent-related mutants, \tool and \botium produced the same results in all chatbots but \texttt{Cloud Assistant}, in which \tool detected two more faults about an intent that \botium could not exercise (79\% and 83\% overall for \botium and \tool, respectively). As expected, both tools missed mutations affecting the fallback mechanism. 

On the other hand, faults in slot types and input/output elements pertaining to flows were more extensively detected by \tool (overall, mutation scores of 58\% and 65\%, respectively), whereas \botium performed poorly (13\% and 43\%, respectively). The performance of \botium can be explained by the simplicity of its generated tests, as already observed for RQ1, since the tool could not thoroughly exercise slot filling mechanism; moreover, by default \botium is limited to the basic intent asserter, which does not specifically check for expected slot values or chatbot responses, hence missing any mutations affecting them. 

Although promising, the outcomes of \tool also suggest some limitations in fault detection. The tool was unable to detect faults affecting flow context data, as the generated tests could not determine their correct expiration timing after such data had been mutated from the original behavior. Further, some mutations affecting removal of prompts for slot filling were not detected, as \tool did not produce tests exercising such conversational paths.

\begin{shadedbox}
    \textbf{Answer to RQ2:} \tool was able to detect more mutants than \botium in all considered groups of mutations and chatbots, achieving a mutation score for slot types four times higher than \botium.
\end{shadedbox}

\subsection{Results for RQ3 - Test Efficiency} 

Table \ref{tab:rq3} shows the time required by \tool and \botium to generate and execute the test suites, respectively, with the corresponding test volume and test complexity (minimum, maximum, and average number of user-bot interactions).

In all cases, the time required for test generation and execution was higher in \tool than in \botium. The test generation time in \botium was very similar to \tool, few seconds in most cases including graph generation. 
On the other hand, \tool required more execution time than \botium. In general, test suites required 4 seconds to 45 minutes for \botium and 3 minutes to 53 minutes for \tool, with a comparable time per test. 

The longer time required by \tool to complete the execution of the test suite was justified by the larger number of tests and the higher complexity of the tests. Regarding complexity, \tool outperformed \botium in every chatbot, producing more sophisticated tests (between 2 and 6 user-bot interactions on average), whereas \botium was limited to generate just a user request and a chatbot response, missing more advanced scenarios. 
For example, considering the \texttt{Cloud Assistant} chatbot, the only subject for which \tool generated fewer tests than \botium, \tool still required more time to complete the test suite, as the higher complexity of its tests enabled conversational scenarios in which the user could create resources through interactions with the chatbot, functionality that could not be exercised by \botium.

\begin{table}[h!]
\caption{Results for RQ3 - Time, Test Volume, and Test Complexity.}
\vspace{-3mm}
\label{tab:rq3}
\centering
\small
\resizebox{\columnwidth}{!}{
\begin{tabular}{|>{\raggedright}p{3.1cm}|c|c|c|c|c|c|}
\cline{2-7}
\multicolumn{1}{c|}{} 
& \multicolumn{3}{>{\columncolor{gray!40}}c|}{\textbf{Botium}} 
& \multicolumn{3}{>{\columncolor{gray!40}}c|}{\textbf{LexTester}} \\
\hline
\rowcolor{gray!40}
\textbf{Chatbot}
& \textbf{Time} & \textbf{\# Tests} & \textbf{Complexity}
& \textbf{Time} & \textbf{\# Tests} & \textbf{Complexity} \\
\hline
Air Line Bot & 4m (0.9s/test) & 249 & 2 & 38m (1s/test) & 2,172 & 1-5 (2 avg.) \\
\hline
Book Trip Manager & 1m (0.7s/test) & 85 & 2 & 13m (2.1s/test) & 375 & 1-10 (5 avg.) \\
\hline
Cloud Assistant & 45m (2.4s/test) & 1,133 & 2 & 53m (5.4s/test) & 585 & 2-3 (3 avg.) \\
\hline
Movie Recommender & 1m (0.8s/test) & 78 & 2 & 3m (1.2s/test) & 146 & 1-6 (3 avg.) \\
\hline
Order Flowers Bot & 4s (2s/test) & 2 & 2 & 15min (2.6s/test) & 349 & 1-9 (6 avg.) \\
\hline
\end{tabular}
}
\vspace{-4mm}
\end{table}

\begin{shadedbox}
    \textbf{Answer to RQ3:} \tool was able to produce more tests in a comparable time than \botium in four cases out of five, exhibiting higher test execution time justified by higher test volume and complexity.
\end{shadedbox}

\subsection{Discussion}

Results across all research questions showed that both \tool and \botium are capable of generating large test suites with low time overhead and comparable execution times. \tool can consistently generate tests of higher complexity, enabling the execution of more diverse and advanced conversational scenarios, leading to broader coverage of conversational and functional aspects than \botium, as also reflected by higher mutation scores. 

While \tool outperforms \botium in all the considered dimensions, it can show limitations in detecting faults affecting the lifespan of flow context data, which would require longer interaction sequences. As previously observed, both tools are limited by their reliance on available training data, which prevents systematic exploration of fallback intents. Although the Dialog Graph generated by \tool explicitly models both successful and failure scenarios, some of the paths are not feasible during test generation, due to the absence of sufficient negative data to properly instantiate them. Finally, despite the high test volume, test redundancy remains an issue. Neither tool was able to fully achieve coverage of all conversational elements together. Nevertheless, \tool reached 100\% coverage of both slot types and slots in four out of \numChatbots cases. 

Overall, these findings are aligned with previous studies on task-based chatbot testing~\cite{ferdinando2024mutabot,rapisarda2025test,gomez2024mutation,clerissi2025towards,bravo2020testing,canizares2024coverage}. At the same time, they highlight the advantages of \tool in exercising complex conversational and functional behaviors,  indicating the need for improved data generation when training data is scarce.
The evaluation was conducted on a heterogeneous set of \numChatbots Amazon Lex chatbots, including two with relatively high conversational complexity compared to other publicly available Lex chatbots, making them representative of non-trivial industrial-like scenarios. Moreover, test generation and execution times remained acceptable even for the more complex chatbots, suggesting that the approach scales well with increasing conversational complexity.

\subsection{Threats to Validity} \label{sec:threats}

An internal threat to validity concerns the manual intervention required during the experiment. In particular, we adapted Mutabot tool to support Amazon Lex chatbots. Further, the assessment of coverage and fault detection required manual inspection of the generated tests and their outcomes, as well as inspection of the produced mutants. 
Third-party chatbots require manual configuration and deployment. 
To mitigate this threat, we carefully inspected each deployed chatbot to check that it worked as intended. Mutants produced by Mutabot were also inspected to ensure that no unintended deviations from the expected mutations occurred. Finally, we inspected test outcomes across multiple test executions, to collect consistent results. 

An external threat to validity concerns the generalizability of the results. Although the findings cannot be considered definitive, they provide an initial empirical assessment of automated testing for Amazon Lex task-based chatbots. The evaluation included both prebuilt and third-party chatbots from different domains and complexities, and results were consistent with previous studies with other platforms~\cite{ferdinando2024mutabot,rapisarda2025test,gomez2024mutation,clerissi2025towards,bravo2020testing,canizares2024coverage}.

%% file: related-work.tex
As chatbots are increasingly integrated into real-world applications, the need for tailored and automated quality assurance techniques, capable of exercising conversational aspects and designing well-defined oracles has arisen. However, only limited approaches have addressed this problem so far~\cite{cabot2021testing,li2022review,lambiase2024motivations,masserini2025brasato}. 

\botium~\cite{botium} is a widely used automated quality assurance framework that supports test generation and execution for task-based chatbots, across a plethora of platforms, both open-source and commercial, including Dialogflow, Rasa, and Amazon Lex.
Over the years, \botium has been used as a backbone for subsequent research works, specifically targeting Rasa and Dialogflow platforms~\cite{bravo2020testing,canizares2024coverage,rapisarda2025test}. CHARM~\cite{bravo2020testing} leverages \botium test cases for test augmentation, generating variants of user utterances aimed at evaluating chatbot robustness. ASYMOB~\cite{canizares2024coverage} generates \botium-like test cases employing multiple coverage strategies. CTG~\cite{rapisarda2025test} uses \botium output as seed tests to automatically augment test suites by capturing dynamic behaviors and improving conversational coverage.

As in traditional approaches, to assess testing effectiveness, mutation testing has been adapted to the chatbot domain, emulating actual faults affecting conversations~\cite{ferdinando2024mutabot,gomez2024mutation,clerissi2025towards,rapisarda2025test}. These studies have revealed a general low capability of existing tools in detecting injected faults. 


Other approaches have explored black-box testing of task-based chatbots through user simulation~\cite{vasconcelos2017bottester,bozic2019chatbot}. For instance, Bottester~\cite{vasconcelos2017bottester} simulates user interactions with chatbots, taking as input specification of conversational flows and time controlled events. Bo{\v{z}}i{\'c} et al.~\cite{bozic2019chatbot} propose an automated testing approach based on AI planning, where a plan represents an abstract test case as actions to perform to achieve a given goal. More recently, TRACER~\cite{del2025automated} infers conversational models from chatbots using Large Language Models (LLMs) and synthesizes them into testing profiles to guide user simulation~\cite{de2025automated}.

Similarly to CHARM~\cite{bravo2020testing}, input mutation for task-based chatbots has been investigated by Guichard et al.~\cite{guichard2019assessing}, introducing paraphrases for evaluating chatbot robustness on the BoTest testing framework~\cite{ruane2018botest}. Input mutation has also been studied concerning the oracle problem, defining metamorphic relations~\cite{chen2018metamorphic}, such as synonym transformations and word removals, and an ontology-based infrastructure to support test generation for Dialogflow chatbots~\cite{bozic2019testing,bovzic2022ontology}.
Dialtest~\cite{liu2021dialtest} was proposed as a testing tool to generate input mutations for evaluating accuracy of NLU modules in recurrent neural network-based dialog systems.

Unlike previous approaches, \tool is the first to enable model-based testing of Amazon Lex chatbots relying on \botium for test execution only.




%% file: conclusion.tex
Task-based chatbots are increasingly being adopted in real-world applications, yet tailored automated testing approaches still struggle to effectively exercise complex conversational and functional behaviors. In particular, research efforts have focused mainly on Dialogflow and Rasa platforms, mostly neglecting Amazon Lex despite being a popular chatbot design platform.

In this paper, we presented \tool, an automated model-based testing approach for Amazon Lex task-based chatbots. \tool was evaluated against the state-of-the-practice \botium tool. Experimental results showed that \tool generates more complex and effective test suites than \botium, improving conversational coverage and fault detection while maintaining comparable execution costs.
As future work, we aim at extending the study of \tool to additional application contexts. At the same time, we plan to enhance the existing \tool functionalities, in particular implementing more advanced coverage criteria, designing input mutations to enable robustness testing and manage scarce training data, studying dynamic and LLM-based oracles, and adapting test suite minimization techniques.